\documentclass[osajnl,twocolumn,showpacs,superscriptaddress,10pt]{revtex4-1} 
\pdfoutput=1
\usepackage{xspace}
\newcommand{\micron}{\,\ensuremath{\mu}{\rm m}\xspace}

\newcommand{\ii}{{\rm i}}

\newcommand{\simlt}{\stackrel{<}{_{\sim}}}

\usepackage{amsmath,amssymb,graphicx}

\begin{document}

\title{Field of view for near-field aperture synthesis imaging}
\author{David F. Buscher}\email{Corresponding author: dfb@mrao.cam.ac.uk}
\affiliation{Cavendish Laboratory, University of Cambridge, J J Thomson Avenue, Cambridge CB3 0HE, UK} 
\begin{abstract}
Aperture synthesis techniques are increasingly being employed to provide high angular resolution images in situations where the object of interest is in the near field of the interferometric array. Previous work has showed that an aperture synthesis array can be refocused on an object in the near field of an array, provided that the object is smaller than the effective Fresnel zone size corresponding to the array-object range. We show here that, under paraxial conditions, standard interferometric techniques can be used to image objects which are substantially larger than this limit. We also note that interferometric self-calibration and phase-closure image reconstruction techniques can be used to achieve near-field refocussing without requiring accurate object range information. We use our results to show that the field of view for high-resolution aperture synthesis imaging of geosynchronous satellites from the ground can be considerably larger than the largest satellites in Earth orbit.
\end{abstract}

\maketitle %% required

%\section{Introduction}
The techniques of aperture synthesis were initially developed for astronomical imaging, where the distance to the object being studied could in most cases be assumed to be infinite. This ``far field'' approximation is not valid for more recent applications such as microwave remote sensing \cite{martin-neira_microwave_2014}, terahertz imaging \cite{bandyopadhyay_terahertz_2006} and tracking of space debris\cite{tingay_detection_2013},  because the objects are sufficiently close to the interferometric array that the curvature of the wavefronts from the target at the location of the imaging array cannot be neglected. 

This ``near field'' condition occurs when the distance $R$ between the array and the object is given by $R\simlt B_{\rm max}^{2}/\lambda$  where $B_{\rm max}$ is the maximum interferometric baseline used and $\lambda$ is the observing wavelength. To give a practical example, obtaining 10\,cm-resolution images from the ground of satellites in geosynchronous orbit requires an interferometer with a maximum baseline length of $B_{\rm max}\sim350$\,m, assuming an operating wavelength of $\lambda=1\,\mu$m. For such an interferometer, near-field effects become important for objects closer than about $10^{8}$\,km, so all geosynchronous satellites, which orbit approximately 36,000\,km from the Earth's surface, are well inside the near field.

Carter \cite{carter1989refocusing} showed that the wavefront curvature occurring in the near field could be compensated for in an interferometer by a procedure analogous to refocussing a telescope. However the analysis assumed that the maximum extent $X_{\rm max}$ of the object satisfies $X_{\rm max}\ll \sqrt{\lambda R}$ where $R$ is the distance from the array to the object, equivalent to the object being smaller than the first Fresnel zone radius for propagation over a distance $R$. This restriction is quite limiting in many cases. In the geosynchronous satellite example given above, only objects much less than 6\,m in size satisfy Carter's assumption. Many satellites in geosynchronous orbit are significantly larger than this, with ``bus'' sizes of 10-15\,m and solar panels up to 50\,m in span.

Lazio \cite{lazio_near-field_2009} discusses a method for getting around this restriction using an adaptation of the ``w-projection'' algorithm used for far-field but wide-angle interferometric measurements. This procedure requires the use of additional data processing steps and cannot directly use standard astronomical image reconstruction software.

In this letter, estimates are derived for the field of view which can be imaged using standard  interferometric image reconstruction software with little or no modification. 
It is shown that objects significantly larger than the field of view suggested by Carter can be straightforwardly imaged in a number of situations of practical interest.

%\section{Imaging model}
\label{sec:model}
\begin{figure}
  \centerline{\includegraphics[width=0.8\columnwidth]{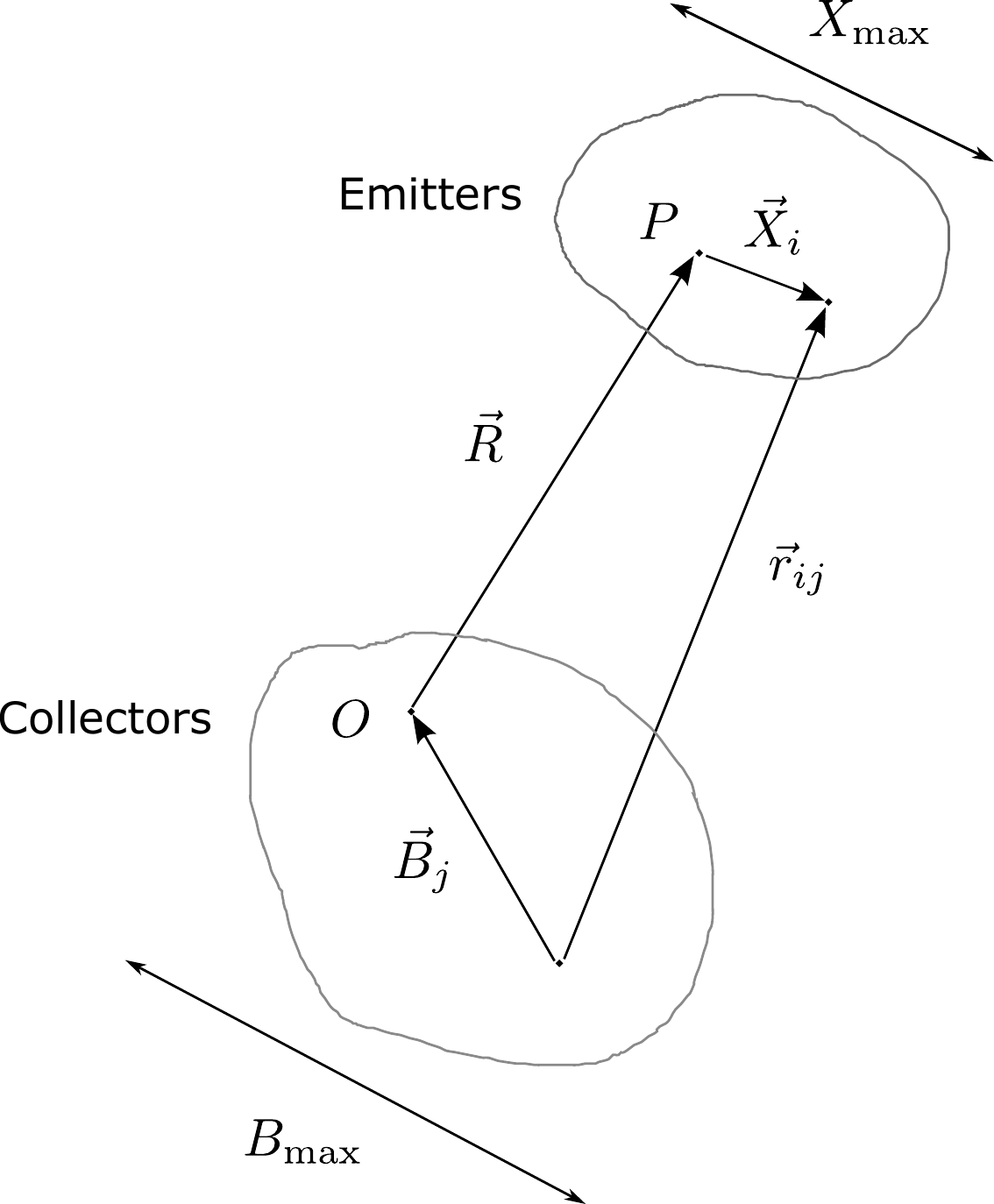}}
  \caption{Geometric model of the interferometer and object.}
  \label{fig:geom-model-interf}
\end{figure}

The geometric model for the imaging system is shown in Figure~\ref{fig:geom-model-interf}. Both the interferometer and the object under observation consist of a set of points in arbitrary three-dimensional arrangements. In the case of the object these points represent the locations of emitters of radiation and in the interferometer these points represent collectors of radiation (telescopes at optical wavelengths and antennae at radio wavelengths). It is assumed that the collectors are clustered around an array center $O$  and that the emitters are clustered around a ``phase center'' $P$. The vector from $O$ to $P$ is denoted by the vector $\vec R$ and its length $R$ is called the ``range''. The choice of the locations $O$ and $P$ is to some extent arbitrary but a good choice for each location is one which minimises the distance to the furthestmost point in the respective cluster.

The vector from $P$ to emitter $i$ is given by $\vec X_i=(x_i, y_i, z_i)R$  where the coordinate system is a Cartesian one such that the $z$ axis is parallel to $\vec{R}$.
The vector from array point $j$ to $O$ is given by $\vec B_j=(u_j,v_j,w_j)R$ using the same coordinate axes.  Note that the $u$, $v$ and $w$ coordinates defined in this way differ from the standard aperture synthesis variables of the same name because they are normalised by the range $R$ rather than by the wavelength $\lambda$.

An important assumption is that the maximum dimension of the object $X_{\rm max }$ and the maximum dimension of the interferometer $B_{\rm max}$ are much less than $R$. 
In other words  $|x_i|^{2}+|y_i|^{2}+|z_i|^{2}\ll 1$ and $|u_j|^{2}+|v_j|^{2}+|w_j|^{2}\ll 1$ for any $i$ or $j$.
This ``paraxial'' condition is more restrictive on the collector array size than the geometry adopted by Carter \cite{carter1989refocusing} who assumed that the collectors can be distributed over a half-sphere surrounding the emitters. However, paraxial conditions are satisfied in many conditions of practical interest, with the geosynchronous satellite example serving as one case in point.

The emitters are assumed to be isotropic emitters and incoherent with one another. Only quasi-monochromatic radiation at a single wavelength $\lambda$ is considered. 
The interferometer is modelled as an abstract device which measures complex ``fringe visibilities'' (more accurately ``coherent fluxes'', since these have units of flux, rather than being normalised to unity for a point source) given by 
\begin{equation}
  \label{eq:4}
  V_{jk}=\left\langle \psi(\vec B_{j}) \psi^{*}(\vec B_{k})\right\rangle,
\end{equation}
where $\psi(\vec B_{j})$ is the instantaneous complex wave amplitude measured at collector $j$ and angle brackets denote averaging over a time much longer than the coherence time of the radiation. In an optical interferometer these visibilities would be obtained by formation of interference fringes and measurement of the fringe parameters, whereas in a radio interferometer the direct product of the measured field amplitudes would be formed and averaged.

%\section{Fringe visibility for a point emitter}
%\label{sec:measvis}

Considering the radiation from a single emitter $i$ of strength $A_{i}$, then using a scalar wave approximation, the complex wave amplitude received at collector $j$ is given by
\begin{equation}
  \label{eq:1}
  \psi(\vec B_{j})\propto\frac{A_{i}\exp\left[2\pi \ii|\vec r_{ij}|/\lambda\right]}{|\vec r_{ij}|}
\end{equation}
where $\vec r_{ij}$ is the vector from collector $j$ to emitter $i$. This vector is given by
\begin{equation}
  \label{eq:2}
  \vec r_{ij}=\vec B_{j}+\vec{R}+\vec X_{i}
\end{equation}
and so writing $\vec{R}=(0,0,1)R$, the distance $r_{ij} = |\vec r_{ij}|$ between the emitter and collector is given by
\begin{equation}
  \label{eq:3}
   r_{ij}  =\left([x_{i}+u_{j}]^{2}+[y_{i}+v_{j}]^{2}+[1+z_i+w_j]^{2}
\right)^{1/2}R.
\end{equation}

Under paraxial assumptions the denominator in equation~(\ref{eq:1}) can be approximated as $r_{ij}\approx R$. Substituting equation~(\ref{eq:1}) into equation~(\ref{eq:4}) gives an expression for the complex fringe visibility
\begin{equation}
  \label{eq:5}
  V_{jk}\propto \frac{|A_{i}|^{2}\exp\left[2\pi \ii d(\vec X_{i},\vec B_{j},\vec B_{k})/\lambda\right]}{R^2}
\end{equation}
where 
$d(\vec X_{i},\vec B_{j},\vec B_{k})$ 
is the optical path difference (OPD) given by
\begin{equation}
  \label{eq:6}
  d(\vec X_{i},\vec B_{j},\vec B_{k})=r_{ij}-r_{ik}.
\end{equation}

Without loss of generality, we can consider the visibility $V_{j0}$ on a baseline where one collector is situated at $O$ and define the OPD as a function of two coordinates
\begin{equation}
  \label{eq:9}
  d(\vec X_{i},\vec B_{j})=d(\vec X_{i},\vec B_{j},0). 
\end{equation}
The following results can be straightforwardly extended to any collector pair in the array, since
\begin{equation}
  \label{eq:10}
  d(\vec X_{i},\vec B_{j},\vec B_{k})=d(\vec X_{i},\vec B_{j})-d(\vec X_{i},\vec B_{k}). 
\end{equation}

Defining a scale parameter $a$ which is the greater of $X_{\rm max}/R$ and $B_{\rm max}/R$, the OPD can be expanded to second order in $a$ to give
\begin{equation}
  \label{eq:8}
  d(\vec X_{i},\vec B_{j})=R \left( x_{i}u_{j}+y_{i}v_{j}+w_j+\frac{1}{2}u_j^2+\frac{1}{2}v_j^{2} 
+\mathcal{O}(a^{3})
\right).
\end{equation}
The first two terms in the expansion can be recognized as giving rise to the linear dependence of the fringe phase on the off-axis angle of the emitter, identical to that for far-field interferometry. The following three terms are independent of the emitter position. The $w$ term is normally removed using an internal delay in the interferometer which  ``points'' the array at the phase center. The quadratic terms arise from the spherical nature of the wavefront from the emitter and so are negligible in the far field when
\begin{equation}
  \label{eq:11}
  R(u_{j}^{2}+v_j^2)\ll \lambda.
\end{equation}
This is equivalent to the standard far-field condition
\begin{equation}
  \label{eq:12}
  R\gg \frac{ B_{\rm max}^{2}}{\lambda}.
\end{equation}

In the near-field, the observed visibility phase can be corrected to the value which would be observed if the source were in the far field by subtracting a phase offset which is quadratically dependent on the distance between the collector and $O$. This correction is equivalent (to second order) to the ``focussing'' correction suggested by Carter \cite{carter1989refocusing}. 
We choose to write this phase correction in terms of the visibility phase which would be observed for a point emitter at the phase centre, so that the phase-corrected visibility is given by
\begin{equation}
  \label{eq:13}
  {V_{j0}}^{\prime}=V_{j0}\exp\left[-2\pi\ii d(0,\vec B_{j})/\lambda\right].
\end{equation}

%\section{Fringe visibility for an extended emitter}
%\label{sec:fringe-visibility-extended}
For a collection of incoherent sources then the phase-corrected visibility will be given by
\begin{equation}
  \label{eq:7}
  V_{j0}^{\prime}\propto \frac{1}{R^2}\sum_{i} |A_{i}|^{2}\exp\left[2\pi \ii d^{\prime}(\vec X_{i},\vec B_{j})/\lambda\right].
\end{equation}
where $d^{\prime}$ is the corrected OPD given by 
\begin{equation}
  \label{eq:15}
 d^{\prime}(\vec X_{i},\vec B_{j})= d(\vec X_{i},\vec B_{j})-d(0,\vec B_{j}).
\end{equation}
If a continuous emitter brightness distribution $I(x,y,z)$ is approximated by a set of discrete emitter brightnesses given by $|A_{i}|^{2}\propto I(x_{i},y_{i},z_{i})\,dxdydz$ then we recover the usual two-dimensional Fourier relationship between the phase-corrected visibility and the (depth-integrated) brightness distribution
\begin{equation}
  \label{eq:16}
  V_{j0}^{\prime}\propto \int \exp\left[2\pi \ii (U_{j}x+V_{j}y)\right]\left( \int I(x,y,z)\,dz \right)\,dxdy,
\end{equation}
where $U_{j}=Ru_{j}/\lambda$ and $V_{j}=Rv_{j}/\lambda$ are the standard interferometric ``u-v'' coordinates. In other words, providing the visibility data is phase-corrected for focus, then images can be reconstructed from near-field interferometric data using standard far-field aperture synthesis software.

It is important to note that explicit phase correction of the visibility data is not required if closure phase or equivalent ``self-calibration'' \cite{Pearson1984} image  reconstruction methods are used. These methods assume that the measured visibility is  corrupted by arbitrary ``antenna-dependent'' phase errors (for example due to atmospheric seeing) and solve for these errors simultaneuosly with solving for the image intensity distribution. It is straightforward to demonstrate that the near-field focus correction as written in equation~(\ref{eq:13}) is in an antenna-dependent form and so the closure phase/self-calibration process will solve for the sum of the near-field phase error and any phase errors due to seeing. Thus if closure phase or self-calibration imaging software is used, then near-field visibility data can be used directly without requiring pre-processing to provide focus correction. 

Using self-calibration or closure phase imaging software to correct the visibility phases has the additional advantage that a precise estimate of the range to the target is not required in order to derive a precise focus correction. For  geosynchronous satellite imaging, the satellite range must be known to better than $0.01$\% in order to compute an a-priori phase correction accurate to 1 radian at 1 micron wavelength, and this kind of range accuracy may be difficult to achieve.

In practice, the high-precision phase correction afforded by self-calibration or closure phase software may need to be accompanied by lower-precision OPD adjustments in hardware in order to deal with temporal coherence effects. For example, the fringes formed in an interferometer operating at a central wavelength of 1\,$\mu$m with a bandpass of 10\,nm will have a fringe envelope which extends over approximately $\pm100\,\mu$m of OPD. When imaging a satellite in geosynchronous orbit using baseline lengths of 350\,m, the OPD error due to near-field effects  will be about 1.7\,mm, and so the instrumental OPD will need to be adjusted compared with the far-field case in order to see high-contrast fringes. This hardware adjustment needs only to be accurate at the sub-100\,$\mu$m level: any remaining phase errors will be antenna-dependent and so can be taken care of by self-calibration or closure phase software.
%%\section{Field of view}
%\label{sec:field-view}

The imaging equation given in equation~(\ref{eq:16}) is correct to second order in the scale parameter $a$. Expanding $d^{\prime}$ to third order in $a$ gives
\begin{equation}
  \label{eq:16b}
    d^{\prime}(\vec X_{i},\vec B_{j})=R \left[ x_{i}u_{j}+y_{i}v_{j}-\epsilon +\mathcal{O}(a^{4}) \right]
\end{equation}
where $\epsilon$ is a third-order error term given by
\begin{equation}
  \label{eq:17}
  \epsilon= ( x_{i}u_j+y_{i}v_{j})(z_{i}+w_{j})
+\frac{1}{2} \,  w_{j} (x_i^{2}+ y_i^{2})
+\frac{1}{2} \,z_{i}(  u_j^{2}+ v_j^{2}).
\end{equation}
It is notable that all the terms in $\epsilon$ contain a factor which depends either on the depth of the array $w_{j}$ or the depth of the object $z_{i}$. This means that if both the array and object are planar so that $w_{j}=z_{i}=0$ for all $i$ and $j$ then the third-order terms vanish and only fourth-order error terms need to be considered.

In general however, the array and/or the object will be three-dimensional, and in this case the field-of-view limitations will arise when the third-order terms lead to phase errors which are comparable to a radian, i.e.\ when $R\epsilon\sim\lambda$. The terms in $\epsilon$ are all products of object-dependent distances and array-dependent distances. If the interferometer is large compared to the object, the largest terms in $\epsilon$ will be of order $B_{\rm max}^{2}X_{\rm max}/R^{3}$, while if the object
is larger than the interferometer, then the largest terms will be of order
$B_{\rm max}X_{\rm max}^{2}/R^{3}$.

In the former case, the third-order phase errors will be negligible when
\begin{equation}
  \label{eq:18}
  X_{\rm max}\ll \left( R\frac{\lambda}{B_{\rm max}} \right) \frac{R}{B_{\rm max}}.
\end{equation}
The factor $R\frac{\lambda}{B_{\rm max}}$ is approximately the size of a resolution element $x_{\rm res}$ in the interferometric image, so the maximum size of object which can be imaged is approximately $R/B_{\rm max}$ resolution elements across. Thus the field of view expression in equation~(\ref{eq:18}) can also be written
\begin{equation}
\label{eq:19}
X_{\rm max}\ll {x_{\rm res}}\left( \frac{x_{\rm res}}{\lambda} \right)
\end{equation}

If instead the object is larger than the interferometer, then the phase errors will be negligible when
\begin{equation}
  \label{eq:20}
  X_{\rm max}^{2}\ll \left( R\frac{\lambda}{B_{\rm max}} \right) R,
\end{equation}
which can be written in terms of the interferometric resolution $x_{\rm res}$ as
\begin{equation}
  \label{eq:21}
  X_{\rm max}\ll \sqrt{{x_{\rm res}}R}.
\end{equation}

%\section{Discussion}
%\label{sec:discussion}
Equations~(\ref{eq:19}) and (\ref{eq:21}) give expressions for the field of view for imaging using standard self-calibration/closure phase image reconstruction methods. It is interesting to apply these results to some cases of practical interest. 

A one-dimensional teraherz aperture synthesis experiment is described by Bandyopadhyay et al.\    \cite{bandyopadhyay_terahertz_2006} using an array of collectors with $B_{\rm max}=5$\,cm and $R=40.9$\,cm. The spatial resolution of this arrangement at the operating frequency of 0.535\,THz (corresponding to $\lambda=0.542$\,mm) is $x_{\rm res}\approx 4.43$\,mm, so using equation~(\ref{eq:19}) gives a value for the field of view of about 8 resolution elements or 36\,mm.  Bandyopadhyay et al.\  image a point source approximately 10 resolution elements away from the optical axis and find the image reconstruction adequate, indicating that the field of view given by equation~(\ref{eq:19}) may be pessimistic in this case.

In the case of a geosynchronous satellite observation aiming for $x_{\rm res}$=10\,cm resolution at $\lambda=1\micron$, then the maximum object size that can be observed is given by equation~(\ref{eq:19}) as 10\,km. This is much larger than the maximum baselines needed  to provide the desired resolution, and so instead equation~(\ref{eq:21}) must be used to estimate the field of view. This gives $X_{\rm max}\approx 1.9$\,km which is still much larger than the 50\,m dimensions of the largest satellite solar panels. 

Geosynchronous satellite imaging at microwave frequencies may be less straightforward: equation~(\ref{eq:19}) suggests that, if 10\,cm resolution is required at a wavelength of 1\,cm, third-order phase aberrations can become become significant for fields of view of as small as 1 meter. However, the field of view improves quadratically as the resolution requirement is relaxed: if 50\,cm resolution is adequate for a given investigation, then imaging a field of view of 25\,m would be unproblematic.

The above analysis shows that standard self-calibration imaging algorithms should work well for geosynchronous satellite imaging applications at visible and near-infrared wavelengths. In this case, factors other than the imaging algorithm, such as providing sufficient dense sampling of the $u-v$ plane \cite{young_interferometric_2013}, are likely to be the key challenges to successful imaging of the largest objects.


\begin{thebibliography}{7}%
\makeatletter
\providecommand \@ifxundefined [1]{%
 \@ifx{#1\undefined}
}%
\providecommand \@ifnum [1]{%
 \ifnum #1\expandafter \@firstoftwo
 \else \expandafter \@secondoftwo
 \fi
}%
\providecommand \@ifx [1]{%
 \ifx #1\expandafter \@firstoftwo
 \else \expandafter \@secondoftwo
 \fi
}%
\providecommand \natexlab [1]{#1}%
\providecommand \enquote  [1]{``#1''}%
\providecommand \bibnamefont  [1]{#1}%
\providecommand \bibfnamefont [1]{#1}%
\providecommand \citenamefont [1]{#1}%
\providecommand \href@noop [0]{\@secondoftwo}%
\providecommand \href [0]{\begingroup \@sanitize@url \@href}%
\providecommand \@href[1]{\@@startlink{#1}\@@href}%
\providecommand \@@href[1]{\endgroup#1\@@endlink}%
\providecommand \@sanitize@url [0]{\catcode `\\12\catcode `\$12\catcode
  `\&12\catcode `\#12\catcode `\^12\catcode `\_12\catcode `\%12\relax}%
\providecommand \@@startlink[1]{}%
\providecommand \@@endlink[0]{}%
\providecommand \url  [0]{\begingroup\@sanitize@url \@url }%
\providecommand \@url [1]{\endgroup\@href {#1}{\urlprefix }}%
\providecommand \urlprefix  [0]{URL }%
\providecommand \Eprint [0]{\href }%
\providecommand \doibase [0]{http://dx.doi.org/}%
\providecommand \selectlanguage [0]{\@gobble}%
\providecommand \bibinfo  [0]{\@secondoftwo}%
\providecommand \bibfield  [0]{\@secondoftwo}%
\providecommand \translation [1]{[#1]}%
\providecommand \BibitemOpen [0]{}%
\providecommand \bibitemStop [0]{}%
\providecommand \bibitemNoStop [0]{.\EOS\space}%
\providecommand \EOS [0]{\spacefactor3000\relax}%
\providecommand \BibitemShut  [1]{\csname bibitem#1\endcsname}%
\let\auto@bib@innerbib\@empty
%</preamble>
\bibitem [{\citenamefont {Mart{\'i}n-Neira}\ \emph {et~al.}(2014)\citenamefont
  {Mart{\'i}n-Neira}, \citenamefont {LeVine}, \citenamefont {Kerr},
  \citenamefont {Skou}, \citenamefont {Peichl}, \citenamefont {Camps},
  \citenamefont {Corbella}, \citenamefont {Hallikainen}, \citenamefont {Font},
  \citenamefont {Wu}, \citenamefont {Mecklenburg},\ and\ \citenamefont
  {Drusch}}]{martin-neira_microwave_2014}%
  \BibitemOpen
  \bibfield  {author} {\bibinfo {author} {\bibfnamefont {M.}~\bibnamefont
  {Mart{\'i}n-Neira}}, \bibinfo {author} {\bibfnamefont {D.~M.}\ \bibnamefont
  {LeVine}}, \bibinfo {author} {\bibfnamefont {Y.}~\bibnamefont {Kerr}},
  \bibinfo {author} {\bibfnamefont {N.}~\bibnamefont {Skou}}, \bibinfo {author}
  {\bibfnamefont {M.}~\bibnamefont {Peichl}}, \bibinfo {author} {\bibfnamefont
  {A.}~\bibnamefont {Camps}}, \bibinfo {author} {\bibfnamefont
  {I.}~\bibnamefont {Corbella}}, \bibinfo {author} {\bibfnamefont
  {M.}~\bibnamefont {Hallikainen}}, \bibinfo {author} {\bibfnamefont
  {J.}~\bibnamefont {Font}}, \bibinfo {author} {\bibfnamefont {J.}~\bibnamefont
  {Wu}}, \bibinfo {author} {\bibfnamefont {S.}~\bibnamefont {Mecklenburg}}, \
  and\ \bibinfo {author} {\bibfnamefont {M.}~\bibnamefont {Drusch}},\ }\href
  {\doibase 10.1002/2013RS005230} {\bibfield  {journal} {\bibinfo  {journal}
  {Radio Science}\ }\textbf {\bibinfo {volume} {49}},\ \bibinfo {pages}
  {2013RS005230} (\bibinfo {year} {2014})}\BibitemShut {NoStop}%
\bibitem [{\citenamefont {Bandyopadhyay}\ \emph {et~al.}(2006)\citenamefont
  {Bandyopadhyay}, \citenamefont {Stepanov}, \citenamefont {Schulkin},
  \citenamefont {Federici}, \citenamefont {Sengupta}, \citenamefont {Gary},
  \citenamefont {Federici}, \citenamefont {Barat}, \citenamefont
  {Michalopoulou},\ and\ \citenamefont
  {Zimdars}}]{bandyopadhyay_terahertz_2006}%
  \BibitemOpen
  \bibfield  {author} {\bibinfo {author} {\bibfnamefont {A.}~\bibnamefont
  {Bandyopadhyay}}, \bibinfo {author} {\bibfnamefont {A.}~\bibnamefont
  {Stepanov}}, \bibinfo {author} {\bibfnamefont {B.}~\bibnamefont {Schulkin}},
  \bibinfo {author} {\bibfnamefont {M.~D.}\ \bibnamefont {Federici}}, \bibinfo
  {author} {\bibfnamefont {A.}~\bibnamefont {Sengupta}}, \bibinfo {author}
  {\bibfnamefont {D.}~\bibnamefont {Gary}}, \bibinfo {author} {\bibfnamefont
  {J.~F.}\ \bibnamefont {Federici}}, \bibinfo {author} {\bibfnamefont
  {R.}~\bibnamefont {Barat}}, \bibinfo {author} {\bibfnamefont {Z.-H.}\
  \bibnamefont {Michalopoulou}}, \ and\ \bibinfo {author} {\bibfnamefont
  {D.}~\bibnamefont {Zimdars}},\ }\href {\doibase 10.1364/JOSAA.23.001168}
  {\bibfield  {journal} {\bibinfo  {journal} {Journal of the Optical Society of
  America A}\ }\textbf {\bibinfo {volume} {23}},\ \bibinfo {pages} {1168}
  (\bibinfo {year} {2006})}\BibitemShut {NoStop}%
\bibitem [{\citenamefont {Tingay}\ \emph {et~al.}(2013)\citenamefont {Tingay},
  \citenamefont {Kaplan}, \citenamefont {McKinley}, \citenamefont {Briggs},
  \citenamefont {Wayth}, \citenamefont {Hurley-Walker}, \citenamefont
  {Kennewell}, \citenamefont {Smith}, \citenamefont {Zhang}, \citenamefont
  {Arcus}, \citenamefont {Bhat}, \citenamefont {Emrich}, \citenamefont {Herne},
  \citenamefont {Kudryavtseva}, \citenamefont {Lynch}, \citenamefont {Ord},
  \citenamefont {Waterson}, \citenamefont {Barnes}, \citenamefont {Bell},
  \citenamefont {Gaensler}, \citenamefont {Lenc}, \citenamefont {Bernardi},
  \citenamefont {Greenhill}, \citenamefont {Kasper}, \citenamefont {Bowman},
  \citenamefont {Jacobs}, \citenamefont {Bunton}, \citenamefont {deSouza},
  \citenamefont {Koenig}, \citenamefont {Pathikulangara}, \citenamefont
  {Stevens}, \citenamefont {Cappallo}, \citenamefont {Corey}, \citenamefont
  {Kincaid}, \citenamefont {Kratzenberg}, \citenamefont {Lonsdale},
  \citenamefont {McWhirter}, \citenamefont {Rogers}, \citenamefont {Salah},
  \citenamefont {Whitney}, \citenamefont {Deshpande}, \citenamefont {Prabu},
  \citenamefont {Shankar}, \citenamefont {Srivani}, \citenamefont
  {Subrahmanyan}, \citenamefont {Ewall-Wice}, \citenamefont {Feng},
  \citenamefont {Goeke}, \citenamefont {Morgan}, \citenamefont {Remillard},
  \citenamefont {Williams}, \citenamefont {Hazelton}, \citenamefont {Morales},
  \citenamefont {Johnston-Hollitt}, \citenamefont {Mitchell}, \citenamefont
  {Procopio}, \citenamefont {Riding}, \citenamefont {Webster}, \citenamefont
  {Wyithe}, \citenamefont {Oberoi}, \citenamefont {Roshi}, \citenamefont
  {Sault},\ and\ \citenamefont {Williams}}]{tingay_detection_2013}%
  \BibitemOpen
  \bibfield  {author} {\bibinfo {author} {\bibfnamefont {S.~J.}\ \bibnamefont
  {Tingay}}, \bibinfo {author} {\bibfnamefont {D.~L.}\ \bibnamefont {Kaplan}},
  \bibinfo {author} {\bibfnamefont {B.}~\bibnamefont {McKinley}}, \bibinfo
  {author} {\bibfnamefont {F.}~\bibnamefont {Briggs}}, \bibinfo {author}
  {\bibfnamefont {R.~B.}\ \bibnamefont {Wayth}}, \bibinfo {author}
  {\bibfnamefont {N.}~\bibnamefont {Hurley-Walker}}, \bibinfo {author}
  {\bibfnamefont {J.}~\bibnamefont {Kennewell}}, \bibinfo {author}
  {\bibfnamefont {C.}~\bibnamefont {Smith}}, \bibinfo {author} {\bibfnamefont
  {K.}~\bibnamefont {Zhang}}, \bibinfo {author} {\bibfnamefont
  {W.}~\bibnamefont {Arcus}}, \bibinfo {author} {\bibfnamefont {N.~D.~R.}\
  \bibnamefont {Bhat}}, \bibinfo {author} {\bibfnamefont {D.}~\bibnamefont
  {Emrich}}, \bibinfo {author} {\bibfnamefont {D.}~\bibnamefont {Herne}},
  \bibinfo {author} {\bibfnamefont {N.}~\bibnamefont {Kudryavtseva}}, \bibinfo
  {author} {\bibfnamefont {M.}~\bibnamefont {Lynch}}, \bibinfo {author}
  {\bibfnamefont {S.~M.}\ \bibnamefont {Ord}}, \bibinfo {author} {\bibfnamefont
  {M.}~\bibnamefont {Waterson}}, \bibinfo {author} {\bibfnamefont {D.~G.}\
  \bibnamefont {Barnes}}, \bibinfo {author} {\bibfnamefont {M.}~\bibnamefont
  {Bell}}, \bibinfo {author} {\bibfnamefont {B.~M.}\ \bibnamefont {Gaensler}},
  \bibinfo {author} {\bibfnamefont {E.}~\bibnamefont {Lenc}}, \bibinfo {author}
  {\bibfnamefont {G.}~\bibnamefont {Bernardi}}, \bibinfo {author}
  {\bibfnamefont {L.~J.}\ \bibnamefont {Greenhill}}, \bibinfo {author}
  {\bibfnamefont {J.~C.}\ \bibnamefont {Kasper}}, \bibinfo {author}
  {\bibfnamefont {J.~D.}\ \bibnamefont {Bowman}}, \bibinfo {author}
  {\bibfnamefont {D.}~\bibnamefont {Jacobs}}, \bibinfo {author} {\bibfnamefont
  {J.~D.}\ \bibnamefont {Bunton}}, \bibinfo {author} {\bibfnamefont
  {L.}~\bibnamefont {deSouza}}, \bibinfo {author} {\bibfnamefont
  {R.}~\bibnamefont {Koenig}}, \bibinfo {author} {\bibfnamefont
  {J.}~\bibnamefont {Pathikulangara}}, \bibinfo {author} {\bibfnamefont
  {J.}~\bibnamefont {Stevens}}, \bibinfo {author} {\bibfnamefont {R.~J.}\
  \bibnamefont {Cappallo}}, \bibinfo {author} {\bibfnamefont {B.~E.}\
  \bibnamefont {Corey}}, \bibinfo {author} {\bibfnamefont {B.~B.}\ \bibnamefont
  {Kincaid}}, \bibinfo {author} {\bibfnamefont {E.}~\bibnamefont
  {Kratzenberg}}, \bibinfo {author} {\bibfnamefont {C.~J.}\ \bibnamefont
  {Lonsdale}}, \bibinfo {author} {\bibfnamefont {S.~R.}\ \bibnamefont
  {McWhirter}}, \bibinfo {author} {\bibfnamefont {A.~E.~E.}\ \bibnamefont
  {Rogers}}, \bibinfo {author} {\bibfnamefont {J.~E.}\ \bibnamefont {Salah}},
  \bibinfo {author} {\bibfnamefont {A.~R.}\ \bibnamefont {Whitney}}, \bibinfo
  {author} {\bibfnamefont {A.}~\bibnamefont {Deshpande}}, \bibinfo {author}
  {\bibfnamefont {T.}~\bibnamefont {Prabu}}, \bibinfo {author} {\bibfnamefont
  {N.~U.}\ \bibnamefont {Shankar}}, \bibinfo {author} {\bibfnamefont {K.~S.}\
  \bibnamefont {Srivani}}, \bibinfo {author} {\bibfnamefont {R.}~\bibnamefont
  {Subrahmanyan}}, \bibinfo {author} {\bibfnamefont {A.}~\bibnamefont
  {Ewall-Wice}}, \bibinfo {author} {\bibfnamefont {L.}~\bibnamefont {Feng}},
  \bibinfo {author} {\bibfnamefont {R.}~\bibnamefont {Goeke}}, \bibinfo
  {author} {\bibfnamefont {E.}~\bibnamefont {Morgan}}, \bibinfo {author}
  {\bibfnamefont {R.~A.}\ \bibnamefont {Remillard}}, \bibinfo {author}
  {\bibfnamefont {C.~L.}\ \bibnamefont {Williams}}, \bibinfo {author}
  {\bibfnamefont {B.~J.}\ \bibnamefont {Hazelton}}, \bibinfo {author}
  {\bibfnamefont {M.~F.}\ \bibnamefont {Morales}}, \bibinfo {author}
  {\bibfnamefont {M.}~\bibnamefont {Johnston-Hollitt}}, \bibinfo {author}
  {\bibfnamefont {D.~A.}\ \bibnamefont {Mitchell}}, \bibinfo {author}
  {\bibfnamefont {P.}~\bibnamefont {Procopio}}, \bibinfo {author}
  {\bibfnamefont {J.}~\bibnamefont {Riding}}, \bibinfo {author} {\bibfnamefont
  {R.~L.}\ \bibnamefont {Webster}}, \bibinfo {author} {\bibfnamefont
  {J.~S.~B.}\ \bibnamefont {Wyithe}}, \bibinfo {author} {\bibfnamefont
  {D.}~\bibnamefont {Oberoi}}, \bibinfo {author} {\bibfnamefont
  {A.}~\bibnamefont {Roshi}}, \bibinfo {author} {\bibfnamefont {R.~J.}\
  \bibnamefont {Sault}}, \ and\ \bibinfo {author} {\bibfnamefont
  {A.}~\bibnamefont {Williams}},\ }\href {\doibase 10.1088/0004-6256/146/4/103}
  {\bibfield  {journal} {\bibinfo  {journal} {The Astronomical Journal}\
  }\textbf {\bibinfo {volume} {146}},\ \bibinfo {pages} {103} (\bibinfo {year}
  {2013})}\BibitemShut {NoStop}%
\bibitem [{\citenamefont {Carter}(1989)}]{carter1989refocusing}%
  \BibitemOpen
  \bibfield  {author} {\bibinfo {author} {\bibfnamefont {W.~H.}\ \bibnamefont
  {Carter}},\ }\href@noop {} {\bibfield  {journal} {\bibinfo  {journal} {IEEE
  Transactions on Antennas and Propagation}\ }\textbf {\bibinfo {volume}
  {37}},\ \bibinfo {pages} {314} (\bibinfo {year} {1989})}\BibitemShut
  {NoStop}%
\bibitem [{\citenamefont {Lazio}(2009)}]{lazio_near-field_2009}%
  \BibitemOpen
  \bibfield  {author} {\bibinfo {author} {\bibfnamefont {J.}~\bibnamefont
  {Lazio}},\ }\href@noop {} {\emph {\bibinfo {title} {On {Near}-{Field}
  w-{Projection} for {Radio} {Interferometric} {Imaging}}}},\ \bibinfo {type}
  {{NRL} {Memorandum} {Report}}\ \bibinfo {number} {NRL/MR/7210--09-9173}\
  (\bibinfo  {institution} {Naval Research Laboratory},\ \bibinfo {year}
  {2009})\BibitemShut {NoStop}%
\bibitem [{\citenamefont {Pearson}\ and\ \citenamefont
  {Readhead}(1984)}]{Pearson1984}%
  \BibitemOpen
  \bibfield  {author} {\bibinfo {author} {\bibfnamefont {T.~J.}\ \bibnamefont
  {Pearson}}\ and\ \bibinfo {author} {\bibfnamefont {A.~C.~S.}\ \bibnamefont
  {Readhead}},\ }\href {\doibase 10.1146/annurev.aa.22.090184.000525}
  {\bibfield  {journal} {\bibinfo  {journal} {Ann. Rev. Astron. Astrophys}\
  }\textbf {\bibinfo {volume} {22}},\ \bibinfo {pages} {97} (\bibinfo {year}
  {1984})}\BibitemShut {NoStop}%
\bibitem [{\citenamefont {Young}\ \emph {et~al.}(2013)\citenamefont {Young},
  \citenamefont {Haniff},\ and\ \citenamefont
  {Buscher}}]{young_interferometric_2013}%
  \BibitemOpen
  \bibfield  {author} {\bibinfo {author} {\bibfnamefont {J.}~\bibnamefont
  {Young}}, \bibinfo {author} {\bibfnamefont {C.}~\bibnamefont {Haniff}}, \
  and\ \bibinfo {author} {\bibfnamefont {D.}~\bibnamefont {Buscher}},\ }in\
  \href {\doibase 10.1109/AERO.2013.6496937} {\emph {\bibinfo {booktitle} {2013
  {IEEE} Aerospace Conference}}}\ (\bibinfo {year} {2013})\ pp.\ \bibinfo
  {pages} {1--9}\BibitemShut {NoStop}%
\end{thebibliography}
\end{document}